\documentclass[twocolumn,showpacs,prl,superscriptaddress]{revtex4}

\usepackage{graphicx}

\usepackage{amssymb}
\usepackage{amsmath}
\usepackage{amsfonts}
\usepackage{mathrsfs}

\renewcommand{\epsilon}{\varepsilon}
\newcommand{\figurewidth}{0.44\textwidth}

\begin{document}
 \title{Dynamical Scaling Exponents for Polymer Translocation through a Nanopore}
\author{Kaifu Luo}
\affiliation{Department of Applied Physics, Helsinki University of
Technology, P.O. Box 1100, FIN-02015 TKK, Espoo, Finland}
\author{Santtu T. T. Ollila}
\affiliation{Department of Applied Physics, Helsinki University of
Technology, P.O. Box 1100, FIN-02015 TKK, Espoo, Finland}
\author{Ilkka Huopaniemi}
\affiliation{Department of Applied Physics, Helsinki University of
Technology, P.O. Box 1100, FIN-02015 TKK, Espoo, Finland}
\author{Tapio Ala-Nissila}
\affiliation{Department of Applied Physics, Helsinki University of
Technology, P.O. Box 1100, FIN-02015 TKK, Espoo, Finland}
\affiliation{Department of Physics, Box 1843, Brown University,
Providence, Rhode Island 02912-1843, USA}
\author{Pawel Pomorski}
\affiliation{Department of Applied Mathematics, The University of Western Ontario,
  London, Ontario, Canada}
\author{Mikko Karttunen}
\affiliation{Department of Applied Mathematics, The University of Western Ontario,
  London, Ontario, Canada}
\author{See-Chen Ying}
\affiliation{Department of Physics, Box 1843, Brown University, Providence,
Rhode Island 02912-1843, USA}
\author{Aniket Bhattacharya}
\affiliation{Department of Physics, University of Central Florida, Orlando,
Florida 32816-2385, USA}

\date{October 14, 2008}
\begin{abstract}

We determine the scaling exponents of polymer translocation (PT)
through a nanopore by extensive computer simulations of various
microscopic models for chain lengths extending up to $N=800$ in some
cases. We focus on the scaling of the average PT time $\tau \sim
N^{\alpha}$ and the mean-square change of the PT coordinate $\langle
s^2(t)\rangle \sim t^\beta$. We find $\alpha=1+2\nu$ and
$\beta=2/\alpha$ for unbiased PT in 2D and 3D. The relation $\alpha
\beta=2$ holds for driven PT in 2D, with crossover from $\alpha
\approx 2\nu$ for short chains to $\alpha \approx 1+\nu$ for long
chains. This crossover is, however, absent in 3D where $\alpha =
1.42 \pm 0.01$ and $\alpha \beta \approx 2.2$ for $N \approx
40-800$.

\end{abstract}
\maketitle

The transport of a polymer through a nanopore plays a crucial role
in numerous biological processes, such as DNA and RNA translocation
across nuclear pores, protein transport through membrane channels,
and virus injection. Due to various potential technological
applications, such as rapid DNA sequencing, gene therapy and
controlled drug delivery, polymer translocation has become a subject
of intensive experimental \cite{Kasianowicz96,Meller03} and
theoretical studies
\cite{Sung,Muthukumar,Chuang,Luo06,Huopaniemi06,Huopaniemi07,Liao,
Milchev,Kantor,Luo062,Luo07,Luo072,Wolterink06,Panja072,Panja07,Panja,Dubbeldam07,
Dubbeldam072,Slater,Storm}.

Among the most fundamental quantities associated with translocation,
the average translocation time $\tau$ as a function of the chain
length $N$ is an important measure of the underlying dynamics.
Standard equilibrium Kramers analysis \cite{Kramers} of diffusion
through an entropic barrier yields $\tau \sim N^2$ for unbiased
translocation and $\tau \sim N$ for driven translocation (assuming
friction is independent of $N$) \cite{Sung,Muthukumar}.
However, as Chuang \textit{et al}. \cite{Chuang} noted, the
quadratic scaling behavior for unbiased translocation cannot be
correct for a self-avoiding polymer. The reason is that the
translocation time is shorter than the Rouse equilibration time of a
self-avoiding polymer, $\tau_R \sim N^{1+2\nu}$, where the Flory
exponent $\nu=0.588$ in 3D and $\nu=0.75$ in 2D \cite{de Gennes},
thus rendering the concept of equilibrium entropy and the ensuing
entropic barrier inappropriate for translocation dynamics. Chuang
\textit{et al}. \cite{Chuang} performed numerical simulations with
Rouse dynamics for a 2D lattice model to study the translocation for
both phantom and self-avoiding polymers. They decoupled the
translocation dynamics from the diffusion dynamics outside the pore
by imposing the restriction that the first monomer, which is
initially placed in the pore, is never allowed to cross back out of
the pore. Their results show that for large $N$, $\tau \sim
N^{1+2\nu}$, which scales approximately in the same manner as the
equilibration time but with a much larger prefactor.
This result was recently corroborated by extensive numerical
simulations based on the Fluctuating Bond (FB) \cite{Luo06} and
Langevin Dynamics (LD) models with the bead-spring approach
\cite{Huopaniemi06,Huopaniemi07,Milchev,Liao}. In Refs.
\cite{Luo06,Huopaniemi06} the translocation time $\tau$ was found to
scale as $N^{2.50 \pm 0.01}$ in 2D, which is in agreement with $\tau
\sim N^{1+2\nu}$.

For driven translocation, Kantor and Kardar \cite{Kantor} have
demonstrated that the assumption of equilibrium in polymer dynamics
by Sung and Park  \cite{Sung} and Muthukumar \cite{Muthukumar}
breaks down more easily and provided a lower bound $\tau \sim
N^{1+\nu}$ for the translocation time by comparison to the unimpeded
motion of the polymer. Using FB \cite{Luo062} and LD
\cite{Huopaniemi06,Luo07} models, a crossover from $\tau \sim
N^{1.46 \pm 0.01} \approx N^{2\nu}$ for relatively short polymers to
$\tau \sim N^{1.70 \pm 0.03} \approx N^{1+\nu}$ for longer chains
was found in 2D.

Recently, however, alternate scaling scenarios have been presented
\cite{Wolterink06,Panja072,Panja07,Panja,Dubbeldam07,Dubbeldam072}, which
contradict the above results.
To resolve the apparent discrepancy, we have undertaken an extensive
effort to determine $\tau$ as function of $N$, $\tau \sim
N^{\alpha}$, and the mean-square change of the translocation
coordinate $\langle s^2(t) \rangle \sim t^\beta$ based on
high-accuracy numerical simulations. The independent models employed
here include the fluctuating bond (FB) model with Monte Carlo (MC)
dynamics \cite{Luo06,Luo062} in 2D, standard Langevin dynamics (LD) of the
bead-spring model of polymers
\cite{Huopaniemi06,Huopaniemi07,Luo07,Luo072} in both 2D and 3D, and
atomistic Molecular Dynamics (MD) simulations using the GROMACS
\cite{van-der-Spoel:05ws} simulation engine in both 2D and 3D.


In the 2D lattice FB model for MC simulation of a self-avoiding
polymer \cite{Luo06,Luo062}, each segment excludes four nearest and
next-nearest-neighbor sites on a square lattice. The bond lengths
$b_l$ vary in length
but do not cross each other. Dynamics is introduced
by Metropolis moves of a single segment, with a probability of
acceptance min[$e^{- U/k_BT}$,1], where $U$ is the energy difference
between the new and old states, $k_B$ the Boltzmann constant and $T$
the absolute temperature. In an MC move, we randomly select a
monomer and attempt to move it onto an adjacent lattice site (in a
randomly selected direction). If the new position does not violate
the excluded-volume or maximal bond-length restrictions, the move is
accepted or rejected according to Metropolis criterion. $N$
elementary moves define one MC time step.

In LD simulations \cite{Huopaniemi07,Luo07,Luo072}, the polymer
chains are modeled as bead-spring chains of Lennard-Jones (LJ)
particles with the Finite Extension Nonlinear Elastic (FENE)
potential. Excluded volume interaction between monomers is modeled
by a short range (SR) repulsive LJ potential with a cutoff of
$2^{1/6}\sigma$, where $\sigma$ is the bead diameter.
Between all monomer-wall particle pairs, there exist the same SR
repulsive LJ interaction. In the LD simulations, each monomer is
subjected to conservative, frictional, and random forces,
respectively.

GROMACS \cite{van-der-Spoel:05ws} is currently one of the most
commonly used programs in soft matter and biophysical simulations,
and has also been used extensively by some of us in various problems
(see e.g., Ref. \onlinecite{Patra:04po} and references therein). As
in the MC and standard LD methods, the hydrodynamic effects are
excluded from our GROMACS MD simulations. The GROMACS MD algorithm
can be implemented with different thermostats. We have used both
overdamped Brownian and Langevin dynamics thermostats. These yield
the same exponents for $\tau$. The results show here and labeled as
MD are from the GROMACS algorithm with LD thermostats.


{\it Unbiased translocation}.
For unbiased translocation, the middle monomer is initially placed
in the center of the pore. The polymer can escape the pore from
either side in time defined as the translocation time $\tau$. We
simulated the escape of chains of lengths varying from $N=15$ up to
$N=255$ for the scaling of $\tau$ and averaged over $200$ samples
for MD simulations in both 2D and 3D and over $2000$ samples for MC
and LD simulations in 2D to minimize statistical errors.

\begin{figure}[tb]
\begin{center}
\includegraphics[width=\figurewidth]{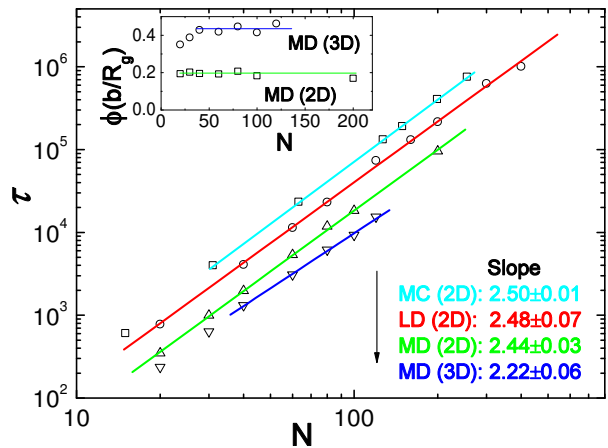}
\caption{Scaling of the translocation time for unbiased case. Curves
have been shifted for clarity. For MC and LD data in 2D, the slopes
are $2.50 \pm 0.01$ \cite{Luo06} and $2.48 \pm 0.07$
\cite{Huopaniemi06}, respectively. For MD data, the slopes are $2.44
\pm 0.03$ (2D) and $2.22 \pm 0.06$ (3D), respectively. The solid
lines indicate fitted data points.} \label{fig1}
\end{center}
\end{figure}

Figure~\ref{fig1} shows $\tau \sim N^{\alpha}$ for different
models. For MD simulations we find that $\alpha=2.44 \pm 0.03$ in
2D and $\alpha=2.22 \pm 0.06$ in 3D, in complete agreement with
$\alpha=2.50 \pm 0.01$ from MC simulations in 2D \cite{Luo06}, and
$\alpha=2.48 \pm 0.07$ from LD simulations in 2D
\cite{Huopaniemi06}. All these results are consistent with the
results from scaling arguments $\tau \sim N^{1+2\nu}$
\cite{Chuang,Luo06,Huopaniemi06} and also agree with the recent
results by Wei \textit{et al.}~\cite{Liao}, where $\tau \sim
N^{2.51 \pm 0.03}$ in 2D and $\tau \sim N^{2.2}$ in 3D based on LD
simulations.

The scaling $\tau \sim N^{1+2\nu}$ implies that $\tau$ scales in the
same manner as the chain equilibration time $\tau_R$. Here, $\tau_R$
is the time it takes a polymer to move a distance equal to its
radius of gyration $\tau_R \sim R_g^2/D$, $D=1/N$ being the
diffusion coefficient. Most recently, Slater {\it et al.}
\cite{Slater} used MD simulations with explicit solvent to study the
impact of hydrodynamic interactions in 3D. The results show that the
scaling of the translocation time varies from $\tau \sim N^{1+2\nu}$
to $\tau \sim N^{3\nu}$ with increasing pore size, which indicates
that the hydrodynamic interaction is screened for small pore sizes.
These results also support $\tau \sim R_g^2/D$ by taking into
account $D \sim 1/N$ and $D \sim 1/N^{\nu}$ without and with
hydrodynamic interactions, respectively. Using a similar argument,
it was also predicted, and numerically confirmed, that $\tau \sim
(R_g+L)^2/D \sim NL^2$ for a long pore of length $L \gg R_g$,
resulting from the fact that the center of mass of the polymer moves
a distance of $L+R_g \approx L$~\cite{Luo06}. For a long pore $L \gg
N$ we have $\tau \sim NL^2 \gg N^3$, which is longer than the
reptation time of the chain $\sim N^3$. In addition, for
translocation under a pulling force $F$ acting on one end of the
chain, $\tau \sim N^{2\nu+1}$ is recovered for $F \rightarrow
0$~\cite{Huopaniemi07}. Altogether, these results further confirm
the argument that $\tau$ scales in the same manner as $\tau_R$.

For the mean-square change of the translocation coordinate $s(t)$,
we use chains of length $N=201$ for LD simulations in 2D and $N=100$
for MD simulations in both 2D and 3D, and average over $2000$
samples. As shown in Fig. 2, we observe sub-diffusive behavior
$\langle s^2(t) \rangle \sim t^{\beta}$, where $\beta=0.80 \pm 0.01$
in 2D for LD simulations and $\beta=0.81 \pm 0.01$ in 2D and
$\beta=0.91 \pm 0.01$ in 3D for MD simulations, as predicted by
Chuang \textit{et al.}~\cite{Chuang}, where
$\beta=2/{\alpha}=2/(1+2\nu)$ gives $0.80$ in 2D and $0.92$ in 3D.

\begin{figure}[tb]
\begin{center}
\includegraphics[width=\figurewidth]{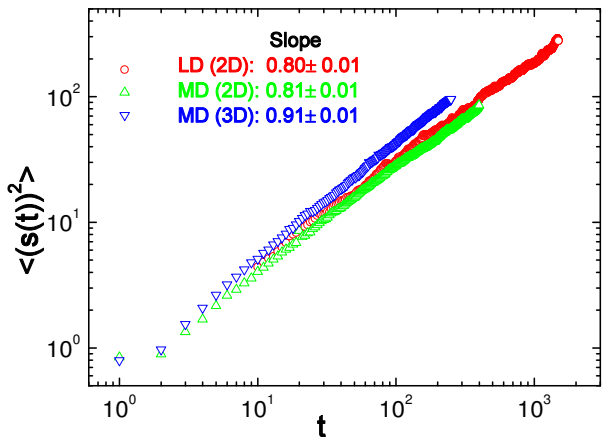}
\caption{Scaling of the mean-square displacement of the
translocation coordinate for unbiased case. For LD data in 2D, the
slope is $0.80 \pm 0.01$ for $10 \le t \le 350$. For GROMACS data,
the slopes are $0.81 \pm 0.01$ (2D) for $10 \le t \le 399$ and $0.91
\pm 0.01$ (3D) for $10 \le t \le 250$, respectively. } \label{fig2}
\end{center}
\end{figure}

All the above results demonstrate that $\alpha=1+2\nu$ and
$\beta=2/(1+2\nu)$ for the range of $N$ studied here. Recently,
Wolterink {\it et al.} \cite{Wolterink06} presented different
results for unbiased translocation using a 3D lattice model with MC
simulations. According to their scaling argument $\tau \sim N^{1 +
2\nu}\phi(b/R_g)$, where $b$ the pore width and the scaling function
$\phi(x) \sim x^{-0.38 \pm 0.08}$ for $x \rightarrow 0$. This leads
to $\tau \sim N^{2.40 \pm 0.08}$ in 3D. For $\tau/N^{1+2\nu}$ as a
function of $N$, as shown in the inset of Fig. 1, we find that the
scaling function $\phi(b/R_g)$ does not depend on $N$, in contrast
to their claims. Furthermore, they have also computed $\langle
s^2(t) \rangle \sim t^{\beta}$, with $\beta=0.81$ in 3D
\cite{Panja072}. In addition, in a more recent paper \cite{Panja07},
two of these authors argue that $\beta=(1+\nu)/(1+2\nu)\approx 0.73$
in 3D for $t<\tau_R$ and it crosses over to $\beta=1$ for
$t>\tau_R$. Correspondingly, they have changed their previous
results to $\tau \sim N^{2+\nu}$, where the exponent is 2.75 in 2D
and 2.588 in 3D.
In the data shown in Fig. 2, there is no sign of such crossover to
$\langle s^2(t) \rangle \sim t$ even at the longest times studied.
In fact, $\tau_R$ is the relaxation time for the whole chain without
confinement. During the translocation process, the chain is always
confined by the pore and thus it is impossible for the whole chain
to be relaxed even if $t>\tau_R$. Therefore, a crossover to the
regime where $\beta=1$ cannot possibly exist.
Most recently, based on the fractional Fokker-Planck equation
Dubbeldam \textit{et al.} \cite{Dubbeldam07} have argued that
$\beta=2/(2\nu+2-\gamma_1)$ and $\tau \sim
N^{2/\beta}=N^{2\nu+2-\gamma_1}$. This gives $\alpha=2.554$ and
$\beta=0.78$ in 2D, where $\gamma_1=0.945$ in 2D, and $\alpha=2.496$
and $\beta=0.80$ in 3D. However, our results disagree with these
claims as well.

{\it Driven translocation}. Various heuristic scaling arguments
for $\tau$ have been presented {\it e.g.} in Refs. \cite{Kantor,
Grosberg, Luo062, Dubbeldam072, Panja}, and will be gauged against
our numerical results below. The simple scaling argument presented
in Ref. \cite{Kantor} gives $1+\nu$ as the upper bound. Using a
more general scaling form with $FR_g/k_BT$ as the relevant
combination \cite{Grosberg} gives that the exponent is bounded
between $2\nu$ and $1+\nu$, while using $FN/k_BT$ gives different
results \cite{Dubbeldam072}. Ref. \cite{Panja} also suggests
$2\nu$ as a lower bound.

%

The translocation time as a function of the polymer length is
presented in Fig. 3. One of the main features is that a crossover
scaling behavior is observed in 2D for using different models. For
short chains $N \le 200$, $\alpha=1.46\pm 0.01$ was found for MC
simulations~\cite{Luo062}, and $\alpha=1.50\pm 0.01$ for LD
simulations~\cite{Huopaniemi06}, and here we find $\alpha=1.52\pm
0.02$ for MD simulations, all of which are consistent with $\tau
\sim N^{2\nu}$. For longer chains, the exponents cross over to
$\alpha=1.70\pm 0.03$ for MC simulations~\cite{Luo062},
$\alpha=1.69\pm 0.04$ for LD simulations~\cite{Huopaniemi06} and
$\alpha=1.64\pm 0.03$ for MD simulations, which are slightly below
the estimate $\tau \sim N^{1+\nu}$.
For simulations in 3D, however, we find no clear evidence of a
crossover predicted by the scaling argument for the range of $N$
studied here. For $N=8-32$ the effective $\alpha$ (running slope) is
close to $2\nu$; however, it rapidly increases with $N$ saturating
to a value of $\alpha={1.42\pm 0.01}$, which is our best estimate
from the new MD data up to $N \le 800$. We note that this value lies
between $2\nu$ and $1+\nu$. Thus, one possible explanation is that
the scaling regime $\tau \sim N^{1+\nu}$ in 3D lies beyond the
values of $N$ studied so far. The MD result is fully consistent with
LD data in 3D where $\alpha={1.41 \pm 0.01}$. As emphasized in the
previous works, driven translocation is a highly non-equilibrium
process \cite{Luo062} and thus simple scaling arguments may not be
accurate. Non-equilibrium effects are expected to be more pronounced
in 3D as compared to the 2D situation. We indeed find that some
aspects of the driven translocation dynamics are sensitive to the
physical system parameters, such as polymer-pore interactions
\cite{Luo072}. Details of these results will be published elsewhere
\cite{Bhatta08}.

In Fig. 4, we show our data for $\langle s^2(t) \rangle \sim
t^{\beta}$, where $\beta=1.36 \pm 0.01$ in 2D and $\beta=1.53 \pm
0.01$ in 3D for LD simulations with $N=128$ and $\beta=1.38 \pm
0.01$ in 2D and $\beta={1.50 \pm 0.01}$ in 3D for MD simulations
with chain lengths $N=100$ and $N=500$, respectively. These
numerical results show that $\alpha \beta=2$ for driven
translocation in 2D. However, in 3D $\alpha \beta \approx 2.2$.

\begin{figure}[tb]
\begin{center}
\includegraphics[width=\figurewidth]{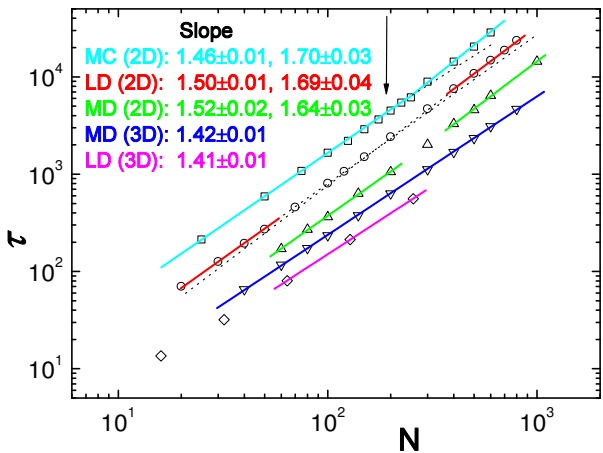}
\caption{Scaling of translocation times under driving force. The
curves have been shifted for clarity. In 2D, the crossover from
$\alpha \approx 2\nu$ to $\alpha \approx 1+\nu$ is observed for all
the simulations. In 3D, there is no clear evidence of such crossover
(see text for details). Solid lines indicated fitted data points.}
\label{fig3}
\end{center}
\end{figure}

Dubbeldam \textit{et al.} \cite{Dubbeldam072} have argued that
$\langle s^2(t) \rangle \sim t^{\beta}$, where
$\beta=4/(2\nu+2-\gamma_1)$. This gives $\beta=1.56$ in 2D and
1.60 in 3D. They further obtain
$\alpha=2/(\beta/2)-1=2\nu+1-\gamma_1$, which gives $\alpha=1.55$
in 2D and $\alpha=1.50$ in 3D. Most recently, Panja \textit{et
al.} \cite{Panja} have argued that $\beta=2(1+\nu)/(1+2\nu)$,
which is 1.40 in 2D and 1.46 in 3D. Numerically, they find that
$\tau \sim N^{(1+2\nu)/(1+\nu)}$, which is 1.43 in 2D and 1.37 in
3D. This result also implies $\alpha \beta=2$. Using the same
argument as Storm \textit{et al.} \cite{Storm}, Panja \textit{et
al.} \cite{Panja} further claim that the lower bound for $\alpha$
is $\tau \sim N^{2\nu}$, which gives $\alpha=1.50$ in 2D.
Obviously, this contradicts both the prediction $\tau \sim
N^{1+\nu}$ in Ref. \cite{Kantor} and the current and previous
simulation results \cite{Luo062,Huopaniemi06,Luo07}.

\begin{figure}[tb]
\begin{center}
\includegraphics[width=\figurewidth]{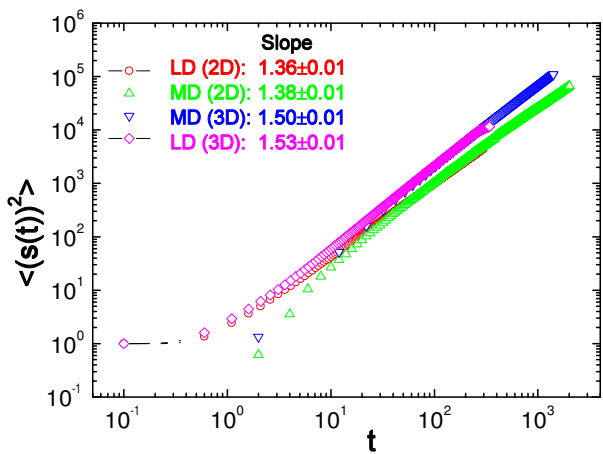}
\caption{ Scaling of the mean-square displacement of the
translocation coordinate for driven translocation. The fitting
regimes are: $10 \le t \le 1500$ (2D LD), $60 \le t \le 2020$ (2D
MD), $40 \le t \le 1416$ (3D MD), and $10 \le t \le 340$ (3D LD).}
\label{fig4}
\end{center}
\end{figure}

To conclude, the dynamics of polymer translocation has been
extensively investigated by several independent models in both 2D
and 3D to conclusively determine the dynamical scaling exponents.
We have focused on the translocation time $\tau$ as function of
the chain length $N$ and the mean-square change of the
translocation coordinate $\langle s^2(t) \rangle \sim t^\beta$.
For unbiased translocation, numerical results are fully consistent
with $\alpha=1+2\nu$ and $\alpha \beta=2$. For driven
translocation, numerical results are again consistent with a
crossover from $\alpha \approx 2\nu$ for short chains to $\alpha
\approx 1+\nu$ in 2D, where the relation $\alpha \beta=2$ is also
valid. In 3D, $2\nu < \alpha=1.42 < 1 + \nu$ and $\alpha \beta
\approx 2.2$ for $40 \le N \le 800$.
These results cast serious doubt on the alternate scaling
scenarios in Refs.
\cite{Wolterink06,Panja07,Panja072,Panja,Dubbeldam07,Dubbeldam072}.

\begin{acknowledgments}
This work has been supported by the Academy of Finland through the
TransPoly and COMP CoE grants, and NSERC of Canada (M.\,K.). We
thank the SharcNet computing facility (www.sharnet.ca) and CSC
Ltd. for computer resources.
\end{acknowledgments}

\end{document}